\documentstyle[12pt]{article}
\begin{document}
\pagenumbering{roman}
\vspace*{0.5in}
\begin{center}
\noindent
{\Large\bf  Persistence Length of Flexible Polyelectrolyte Chains}     

\vspace{.25in}

 B.-Y. Ha$^{(1)}$ and D. Thirumalai$^{(2)}$\\
{\it ${}^1$Department of Chemistry and Biochemistry \\
University of California at Los Angeles \\
Los Angeles, CA 90065 \\
 ${}^2$Institute for Physical Science and Technology\\
University of Maryland\\
College Park, MD 20742}\\

\vspace{1cm}

{\bf ABSTRACT}
\end{center}
We calculate the dependence of the electrostatic persistence length,
$l_e$, of weakly charged flexible polyelectrolyte chains using a
self-consistent variational theory.  The variation of $l_e$ with
$\kappa$, the inverse Debye screening length, is controlled by the
parameter $l_0 l_B/A^2$ where $l_0$ is the bare persistence length,
$l_B$ is the Bjerrum length, and $A$ is the mean distance between
charges along the chain.  Several distinct regimes for the 
dependence of $l_e$ on
$\kappa$ emerge depending on the value of $l_0 l_B/A^2$.  
We show that when $l_0 l_B /A^2 \ll 1$ we
recover the classical result, $l_e \propto \kappa^{-2}$.  For
intermediate values of $l_0 l_B /A^2$, $l_e \propto \kappa^{-1}$.  In
this regime one can also get $l_e \propto \kappa^{-y}$ with $y<1$
depending on the strength of the Coulomb interaction.  Qualitative
comparisons between our theory and simulations as well as other
theories are presented.

\noindent
PACS number(s): 36.20.-r, 33.15.Bh, 64.60.-i, 41.20.-q

\newpage
\pagenumbering{arabic}
\section{Introduction}

In recent years, significant attention has been paid to the old
problem of the ionic dependence of the size of isolated
polyelectrolyte chains [1-4].  The size of the chain is 
most conveniently expressed in terms of the electrostatic persistence
length.  The computation of the persistence length of a
polyelectrolyte chain is difficult because of the interplay of several
length scales.  Significant steps in the computation of the size of
charged polymers was taken by Odjik~\cite{odjik} and Skolnick and
Fixman~\cite{fixman} (OSF), who
introduced the concept of the electrostatic persistence length.  These
authors considered a stiff polyelectrolyte chain near the rod limit so
that only small excursions from linear conformations are relevant.  By
assuming that charges on the rod-like polyelectrolyte interact via the
Debye H\"uckel potential, ${v(r) \over k_BT}= l_B {\rm
e}^{-\kappa r}/r$, where $l_B=e^2/4 \pi \epsilon_0 k_B T$ is the
Bjerrum length, $A$ is the average distance between charges,
$\kappa^{-1}$ is the screening length, OSF showed that the total
persistence length of the chains can be written as $l_p=l_0+l_e$.  Here $l_0$ is the bare
persistence length, and the electrostatic persistence length $l_e$
 in the OSF theory is
$$
l_{OSF}={1 \over 4 } {l_B \over \kappa^2 A^2}
\eqno(1.1) 
$$
It is believed that as long as the chain is stiff the dependence of
$l_e$ on $\kappa$ 
given above is correct [2]. The above result is exact in large dimensions
regardless of the intrinsic stiffness of the chain backbone.

Many experiments on polyelectrolytes have been interpreted
using the OSF theory.  However, light scattering experiments on some
polyelectrolytes systems do show deviations from the OSF theory~\cite{forster,tricot,piazza}.  The
major approximation in the OSF treatment is that the polyelectrolyte
is rod-like, and hence deviations from linear configuration is small.
 If the intrinsic backbone of the polymer chain is flexible, so that
large excursions from linearity are possible, it is likely that
 the OSF result may not be valid.  In fact, the validity of Eq. (1.1) for
chains that are intrinsically flexible has been questioned
using theoretical arguments [7-9] and computer simulations~\cite{
micka,stevens}.  Starting with
the work of Barrat and Joanny~\cite{joanny} (BJ), several variational  
calculations~\cite{thirum.ma,dawson} show
that, when the chains are intrinsically flexible, the electrostatic
persistence length $l_e \sim \kappa^{-1}$.  We had shown
earlier~\cite{thirum.ma} that the predicted 
$\kappa^{-1}$ is valid when the controlling parameter $l_0
l_B/A^2 \sim {\cal O}(1)$, i.e., when the chain is in the so called 
non-asymptotic regime~\cite{thirum.ma}.  Li and
Witten~\cite{witten} argued that, unlike the variational theories~\cite{thirum.ma,joanny,dawson}, approximate
inclusion of fluctuations still leads to the result predicted by OSF
even if the backbone is flexible.  In other words, the OSF result
given in Eq. (1.1) is always valid.  In the above referenced
variational theories it is generally believed that the electrostatic
interactions generally stiffen the chain significantly so that the trial
Hamiltonian should consist of terms that account for the
electrostatic-mediated interaction 
rigidity.  In contrast, Muthukumar has examined the influence of
electrostatic interactions on the size of the flexible polyelectrolyte
chains~\cite{muth}.  This theory, which also emphasize the role of
excluded volume interactions, has been used to interpret
experimental data~\cite{beer}.  A precise form of the ionic dependence of
the electrostatic persistence length that covers all values of
stiffness has not emerged from any these
studies. 

More recently, Micka and Kremer~\cite{micka} have extracted the persistence length
of weakly charged polyelectrolytes of varying stiffness using Monte
Carlo simulations.  They argue that there is no apparent simple scaling of
$l_e$ with $\kappa$ in all regimes.  This raises the possibility that a simple
concept of electrostatic persistence length may not be useful in
determining the size of polyelectrolyte chains under diverse conditions.  The numerical results
can be apparently be rationalized using a field-theoretical renormalization
group analysis of flexible chains.  Liverpool and Stapper~\cite{liverpool} suggest that
the scaling form for the mean square end-to-end distance, $\left< R^2
\right>$, provides a good  fit of the computer simulation data.

The preceding survey suggests that we still do not have a simple theory for
computing $l_e$ for charged polymer chains with arbitrary stiffness.  The
purpose of this paper is  to reexamine the variational type theories
for computing  $l_e$ using the trial Hamiltonian suggested by BJ for flexible chains~\cite{joanny}.  Rather than compute the approximate free
energy using the Gibbs-Bogoliubov inequality as BJ did, we calculate $\left< R^2
\right>$ directly using the uniform expansion method~\cite{es}.  We show
explicitly that when $l_0 l_B/A^2 \ll 1$, the OSF result given in
Eq. (1.1) is recovered.  The rest of the paper is organized as follows.
 In Sec. 2, we present the variational theory for the electrostatic
persistence length in the asymptotic limit, i.e., when $l_0 l_B /A^2
\ll 1$.  The case when $l_0 l_B \sim {\cal O}(1)$, which is most
appropriate for examining the crossover regime, is presented in
Sec. 3.  Comparisons of the theoretical predictions to the simulation
result are discussed in Sec. 4.  The paper is concluded in Sec. 5 with
a few 
additional remarks.

\section{Variational Theory: Chain under tension} 

The flexible polyelectrolyte chain consists of $N$ segments each of
length $l_0$ on an average.  The charges on the chain are assumed to
be a distance $A$ apart.   The interaction
between the charges on the chain is taken to be given by the
Debye-H$\ddot{\rm u}$ckel (DH) potential.  In this model the
counterions are not 
explicitly treated.  In the continuum limit the
partition function of the polyelectrolyte chain is taken to be 
$$
Z=\int {\cal D}[{\bf r}(s)]{\cal W}[{\bf r}(s)]
\eqno(2.1a)
$$
where the statistical weight associated with a given realization of
the polyion described in terms of the curve ${\bf r}(s)$ is 
$$
{\cal W}[{\bf r}(s)]={\cal W}^{\circ}[{\bf r}(s)] {\rm exp} \left[-{1
\over 2}
{l_{\rm B}\over A^2}  \int_0^L\!\! \int_0^L {\rm d}s {\rm d}s'  
  {{\rm e}^{-\kappa \vert {\bf r}(s')-{\bf r}(s) \vert} 
  \over {\vert {\bf r}(s')-{\bf r}(s) \vert}} \right], 
\eqno(2.1b)
$$
and ${\cal W}^{\circ}[{\bf r}(s)]$ is the weight for the chain
backbone in the absence of charges
$$
{\cal W}^{\circ}[{\bf r}(s)]={\rm exp}\left[-{3 \over 2 l} \int_0^L 
\left({\partial {\bf r}(s) \over  \partial s}\right)^{2} \right]
\eqno(2.1c)
$$
with $l=2l_0$.  In order to eliminate the self-interaction in the DH
potential
we impose the condition that $|s'-s| > s_0$ where $s_0 \sim l_0$.
The cut-off dependence vanishes in the variational theory described
below.

The direct evaluation of $Z$ in Eq. (2.1a) is not possible due to the 
non-Markovian nature of the electrostatic interaction term.
Consequently, we have to resort to an approximate calculation.  Here we
use a variational method, first proposed by Edwards and
Singh~\cite{es} (ES),
 in their
treatment of the standard excluded volume problem in neutral flexible polymers.
The
basic idea of the ES method is to replace the weight by a simpler
trial weight with undetermined parameters.  The parameters are chosen
so that the size (say the mean square end-to-end distance) calculated
using the trial weight, $\bigl< R^2 \bigr>_{\rm t}$ becomes identical
to that given by the true weight, $\bigl< R^2 \bigr>$.  The practical
implementation of this method requires that the interaction term can
somehow be treated as small so that the deviation between
$\bigl< R^2 \bigr>_{\rm t}$ and $\bigl< R^2 \bigr>$ can be computed in
powers of the small parameter.
The ES method has been successfully used to predict the conformational
properties of flexible polyions under a variety of solvent 
conditions~\cite{muth,thirum.pra}.

A word of caution concerning the ES method (and in general variational
methods) is in order.  Firstly, the results of variational calculations
are only as good as the chosen trial Hamiltonian.  In polymer problems
there is an additional distinction.  ES advocate that the variational
parameters be chosen to ensure $\bigl< R^2 \bigr>_{\rm t}= \bigl< R^2
\bigr>$
rather than by employing the Gibbs-Bogoliubov inequality, i.e., by
minimizing the free energy.  For polymers with short range
interactions it has been shown that the ES method is superior to 
others that minimize the free energy by a suitable variational calculation.
For 
problems with
long range interactions the direct minimization of free energy can give
reliable results for the scaling of $\bigl< R^2 \bigr>$ with $N$.  For
the present problem the range of interaction is neither short nor 
too large.  From this discussion it should be clear that the utility
of the variational methods should be viewed with some caution.

The choice of the trial weight or Hamiltonian is extremely important.
An appropriate trial Hamiltonian is chosen by considering the case of
$\kappa=0$.  Here, the weakly charged flexible polyelectrolyte can be
considered as an array of blobs of size $D$.  The structure of the
chain inside a blob resembles that of a non-interacting (i.e., the
electrostatic interactions are unimportant) while there is ordering
among the blobs due to the Coulomb repulsion beyond the length scale
$D$.  The physical picture of the charged chain is that of one under
tension~\cite{degennes}, where the blob size (short length scale effect) is determined
by the magnitude of the (as yet undetermined value of the tension), while at long
length scales the chain is a free chain with an effective step length
different from $l_0$.  Thus, in order to determine the structure of
such polyions, it is crucial to incorporate the presence of two length
scales, namely, $D$ that describes the local ordering and
correlations between the semiflexible blobs that
gives the effective persistence length.  Based on this physically 
appealing picture, Barrat and Joanny~\cite{joanny} proposed a
trial probability weight ${\cal W}_{\rm t}[{\bf r}(s)]$ associated
with the configuration $[{\bf r}(s)]$ that succinctly captures the 
conformation of weakly charged polyions described above.  In particular they chose
$${\cal W}_{\rm t}[{\bf r}(s)] = {\cal W}^{\circ}[{\bf r}(s)]
                   \int {\cal D}[{\bf t}(s)] {\cal W}[{\bf t}(s)]
                   {\rm exp} \left[{\tau} \int _0^L {\rm d}s {\bf t}(s) 
                   \cdot {{\partial {\bf r}(s)} \over {\partial s}}
\right], 
\eqno(2.2)
$$
where ${\cal W}^{\circ}[{\bf r}(s)]$ is the Boltzmann weight for the
non-interacting chain (Cf Eq. (2.1c)), ${\bf t}(s)$ is the unit vector
that gives the direction of the tension at position $s$.  The weight 
${\cal W}_{\rm t}[{\bf r}(s)]$ is associated with a particular
realization of the direction of the chain tension and is given by
$$
{\cal W}[{\bf t}(s)] =\prod_{0 \le s \le L}\delta({\bf t}(s)^2-1)\ {\rm
exp} 
\left[-{\eta
\over 2} \int_0^L {\rm d}s  
                           \left({\partial {\bf t}(s) \over \partial
s} \right)^2 \right]
.\eqno(2.3)
$$
The weight associated with the tension vector ${\bf t}(s)$ is an attempt
to capture the physics that the charged chain on a scale $\eta$
 behaves as one under tension.  On the scale $\eta$ the chain
is adequately stiff so that it can be described as worm-like for which
the Boltzmann weight is approximately given by Eq. (2.3).  The finite
range of correlations among the blobs is described by the correlation
in ${\bf t}(s)$ such that 
$$ {\bigl<{\bf t}(s) \cdot {\bf t}(s') \bigr>_{\cal W} = {\rm
exp}[-{\vert s'-s \vert}/{\eta}]}
.\eqno(2.4)
$$
The trial weight has two parameters namely $\eta$ (the effective
persistence length) and the chain tension $\tau$.  Both these are
adjusted so that the deviation between $\bigl< R^2 \bigr>$ and $\bigl<
R^2 \bigr>_{\rm t}$ is zero to first order in the interaction (second
term in Eq. (2.1b)) term. 

From the form of the trial weight ${\cal W}_{\rm t}[{\bf r}(s)]$ it is
clear that the blob size $D \simeq \tau^{-1}$.  This can be formally
established by computing $ \bigl<{\vert {\bf r}(s')-{\bf r}(s)
\vert}^2 \bigr>_{{\cal W}_{\rm t}}$ which can be obtained from the
generating functional defined by
$$
{\cal L}({\bf k};s',s)=  
 \left< {\rm e}^{i {\bf k}
\cdot ({\bf r}(s')-{\bf r}(s))} \right>_{{\cal W}_{\rm t}}
.\eqno(2.5)
$$
The evaluation of ${\cal L}({\bf k};s',s)$ up to order ${ k}^2$ is
given by 
$$
{\cal L}({\bf k};s',s)={\rm e}^{-k^2 l |s'-s|/6}\ \left< {\rm e}^{i {\bf
k} l {\tau}/3 \cdot \int_s^{s'} {\bf t}(s''){\rm d}s''} \right>_{{\cal
W}_{t}}  \qquad \qquad \qquad \qquad \qquad \ $$
$$
\qquad \qquad \quad =-\mbox{\small$\frac{1}{6}$}k^2 l |s'-s|\! -
\mbox{\small$\frac{1}{18}$} l^2
\tau^2  \int_s^{s'}\!\!\int_s^{s'} {\rm d}s_1 {\rm d}s_2 \left<
{\bf k}\cdot {\bf t}(s_1) \ {\bf k} \cdot {\bf t}(s_2) \right>_{{\cal 
W}_{t}}
 $$
$$
\qquad \qquad =-\!\mbox{\small$\frac{1}{6}$}k^2 l |s'-s|\!- 
\mbox{\small$\frac{1}{54}$}k^2 l^2 \tau^2 
 \int_s^{s'}\!\!\int_s^{s'} {\rm d}s_1 {\rm d}s_2 \left< {\bf
t}(s_1) \cdot {\bf t}(s_2) \right>_{{\cal W}_{t}}
.\eqno(2.6)
$$
The mean square internal distance is 
$$
\bigl<{\vert {\bf r}(s')-{\bf r}(s) \vert}^2 \bigr>_{{\cal W}_{\rm t}}
=-6(\partial/\partial
k^2)|_{{\bf k}=0}{\cal L}({\bf k};s',s) \qquad\qquad\qquad\qquad\qquad $$
$$\qquad\qquad\qquad\qquad={\vert s'- s \vert} +
\mbox{\small$\frac{1}{9}$}l^2 {\tau}^2  [2{\eta} {\vert s'-s \vert} 
  - 2{\eta}^2 (1-{\rm e}^{- {\vert s'-s \vert }/{\eta}})],
\eqno(2.7)
$$
where we have used Eq. (2.4).  By equating the two terms in Eq. (2.7)
we can infer the contour length $S \simeq 1/l \tau^2$ and the
blob size $D \sim (S l)^{1/2} \simeq \tau^{-1}$.  The blob size is
naturally expressed also in terms of $l_0$, $l_{\rm B}$, and $A$ and
is given by Eq. (2.7).  For long chains i.e., $L \gg S$ we can neglect
the first term in Eq. (3.7), and the size of the chain $\bigl< R^2
\bigr>_{{\cal W}_{\rm t}}$ becomes 
$$
\bigl<R^2 \bigr>_{{\cal W}_{\rm t}} = \bigl<{\vert {\bf r}(L)-{\bf r}(0)
\vert}^2 \bigr>_{{\cal W}_{\rm t}}
                      =2l_{\rm p} L' - 2l_{\rm p}^2(1-{\rm e}^{- L'/l_{\rm
p}})
,\eqno(2.8) 
$$
where $l_{\rm p} \simeq \eta l \tau$ is the physically observable
persistence length and $L'$ is the longitudinal size of the
corresponding stretched chain
$$
L'\approx l D^{-1}L \approx (l_0l_{\rm
B}/A^2)^{1/3} L
.\eqno(2.9)
$$
Note that the notion of stretched blobs is only valid for $l_{\rm p}/D
\gg 1$.  

In order to set up the calculation of $l_{\rm p}$ using the ES method
we write
$$\bigl< R^2 \bigr>=\bigl< R^2 \bigr>_{{\cal W}_{\rm t}}
      +\Delta R^2 (l_{\rm p};l_{\rm B},A,\kappa)
\eqno(2.10)
$$
where $\bigl< R^2 \bigr>$ is to be computed using the weight given in
Eq. (2.1).  As explained before the variational parameters $\eta$ and
$\tau$ are determined from the condition $\Delta R^2 \equiv 0$.  To
obtain an explicit expression for $\Delta R^2$ let us write 
$$
\bigl<R^2 \bigr>={\int {\cal D}[{\bf r}(s)] \ R^2 \ {\cal W}[{\bf r}(s)]
\over \int {\cal D}[{\bf r}(s)] \ {\cal W}[{\bf r}(s)]}
\qquad\qquad\qquad $$
$$
\qquad\qquad={\int {\cal D}[{\bf r}(s)] \ {\cal W}_{\rm t}[{\bf r}(s)] \
R^2 \ {\rm
exp} \{-B[{\bf r}(s)] \} \over \int {\cal D}[{\bf r}(s)] \ {\cal W}_{\rm
t}[{\bf r}(s)] \ {\rm exp} \{-B[{\bf r}(s)] \} }
\eqno(2.11a)$$
where 
$$
B[{\bf r}(s)]={ l_{\rm B} \over 2 A^2} \int_0^L\!\int_0^L {\rm d}s {\rm
d}s'{{\rm e}^{-\kappa|{\bf r}(s')-{\bf r}(s)|} \over |{\bf r}(s')-{\bf
r}(s)|}+ \tau \int_0^L {\rm d}s {\bf t}(s) \cdot {\partial {\bf r}
\over \partial s}  -{\eta \over 2} \int_0^L {\rm d}s \left({\partial
{\bf t} \over \partial s} \right)^2
.\eqno(2.11b) $$
If ${\cal W}_{\rm t}[{\bf r}(s)]$ is a good approximation to ${\cal
W}[{\bf r}(s)]$ then $B[{\bf r}(s)]$ can be treated as small.  
Consequently up to ${\cal O}(B[{\bf r}(s)])$ the
vanishing of $\Delta R^2$ leads to the condition:
$$
{\bigl< R^2 \ B[{\bf 
r}(s)]  \bigr>_{{\cal W}_{\rm t}}= \bigl<R^2 \bigr>_{{\cal W}_{\rm
t}} \bigl< B[{\bf r}(s)] \bigr>_{{\cal W}_{\rm t}} }
.\eqno(2.12)
$$
By using the following equations
$$
\eta {\partial \over \partial \eta} \bigl< R^2 \bigr>_{{\rm
t}}
=
\bigl< R^2  \bigl>_{{\cal W}_{\rm t}} 
\Bigl< {\eta \over 2} \int {\rm d} s \Bigl({\partial {\bf t} \over
\partial s}
\Bigr)^2   \Bigr>_{{\cal W}_{\rm t}}  
-\left< R^2 \ {\eta \over 2} \int {\rm d} s \left({\partial {\bf t}
\over \partial s}
\right)^2   \right>_{{\cal W}_{\rm t}}
\eqno(2.13a)
$$
$$\tau {\partial \over \partial \tau} \bigl< R^2 \bigr>_{{\cal W}_{\rm
t}}=
-
\bigl< R^2  \bigl>_{{\cal W}_{\rm t}} 
\left< \tau \int {\rm d} s {\bf t}(s) \cdot {\partial {\bf r} \over
\partial s}
  \right>_{{\cal W}_{\rm t}}  
+\left< R^2 \ \tau \int {\rm d} s {\bf t}(s) \cdot {\partial {\bf r}(s)
\over \partial s}
  \right>_{{\cal W}_{\rm t}}
.\eqno(2.13b)
$$
eq. (2.12) may be written as
$$
l_{\rm p} 
{\partial \over \partial l_{\rm p}} \bigl<R^2\bigr>_{{\cal W}_{\rm t}}-
{\tau} {\partial  \over  \partial {\tau}}\bigl<R^2\bigr>_{{\cal W}_{\rm
t}}
=\bigl<R^2 V\bigr>_{{\cal W}_{\rm t}}-
\bigl<R^2\bigr>_{{\cal W}_{\rm t}}\bigl<V\bigr>_{{\cal W}_{\rm t}}
\eqno(2.14)
$$
where $V$ is the total interaction energy given by the first term in
Eq. (2.11b).  Using Eq. (2.8) for $\bigl< R^2 \bigr>_{{\cal W}_{\rm
t}}$ the L.H.S. of Eq. (2.14) can be easily calculated.

The computation of $\bigl< R^2 V\bigr>_{{\cal W}_{\rm t}}$ and $\bigl<
V\bigr>_{{\cal W}_{\rm t}}$ is somewhat involved.  Here we provide of a
sketch
of the calculation.  The details are given in Appendix A.  An
expression of $\bigl< R^2 V\bigr>_{{\cal W}_{\rm t}}$ can be obtained
in terms of the generating functional 
$$
{\cal L}_1({\bf k};0,L|{\bf q};s,s') =\Bigl<{\rm e}^{
i{\bf k}\cdot ({\bf r}(L)-{\bf r}(0))+i{\bf q}\cdot ({\bf r}(s')-{\bf
r}(s))
}\Bigr>_{{\cal W}_{\rm t}} 
\eqno(2.15)
$$
with the result
$$
\bigl< R^2 V \bigr>_{{\cal W}_{\rm t}}
={l_{\rm B} \over 2 A^2} \int\!\int {\rm d}s {\rm d}s' \int {{\rm
d}^3q \over (2 \pi)^3}{4 \pi \over q^2 + \kappa^2}
 \Bigl[-6 {\partial \over \partial {\bf k}^2} \Bigr]_{{\bf k}=0}{\cal
L}({\bf k};0,L|{\bf q};s,s')
.\eqno(2.16)
$$
Similarly
$$
\bigl<R^2\bigr>_{{\cal W}_{\rm t}}\bigl<V\bigr>_{{\cal W}_{\rm t}}
= \left[-6 {\partial \over \partial {\bf k}^2} \right]_{{\bf k}=0}
{\cal L}({\bf k};0,L)
{l_{\rm B} \over 2 A^2} \int\!\int {\rm d}s {\rm d}s' \int {{\rm
d}^3q \over (2 \pi)^3}{4 \pi \over q^2 + \kappa^2}{\cal L}({\bf q};s,s')
\eqno(2.17)
$$
where ${\cal L}({\bf q};s,s')$ is given by Eq. (2.5).

In order to obtain a tractable expression for $l_{\rm p}$ we make an
additional, physically motivated, approximation to evaluate Eqs. (2.16)
and (2.17).  On a scale larger than $\eta$ the chain is effectively
Gaussian provided $L/ \eta \gg 1$.  Thus, the effect of stiffness,
arising from stretching due to electrostatic repulsion, is important
only on the scale of the order $\eta$.  The chain on the scale of
$\eta$ can be treated as essentially rod-like with minimal 
conformational fluctuations.  In order to obtain the generating
functional ${\cal L}$ and ${\cal L}_1$ we need to compute terms with the
structure $
\Bigl<{\rm e}^{i{\bf Q} \cdot \int_0^L {\bf t}(s){\rm d}s}
\Bigr>_{{\cal W}_{\rm t}} $.  With the approximation that on scale of
the order of $|s'-s|=\eta$ the chain behaves essentially as rod-like we
can write
$$
\biggl<{\rm e}^{i{\bf Q} \cdot \int_0^L {\bf t}(s){\rm d}s}
\biggr>_{{\cal W}_{\rm t}}
\approx \biggl<{\rm e}^{i{\bf Q} \cdot \bigl(\int_0^s +\int_{s'}^L \bigr)
{\bf
t}(s){\rm d}s} \biggr>_{{\cal W}_{\rm t}}
\biggl<{\rm e}^{i{\bf Q} \cdot \int_s^{s'}  {\bf
t}(s){\rm d}s} \biggr>_{\rm rod}
.\eqno(2.18)
$$
With this physically motivated approximation the generating functional
can be calculated.  The result for the persistence length can be
written as (after considerable algebra)
$$l_{\rm p} \approx {l_{\rm B} \over A^2}{1 \over L'} 
  {\int\!\!\int} {\rm d}s {\rm d}s'
 \int {{{\rm d}^3q } \over {q^2+{\kappa}^2}}
   {\rm e}^{-{q^2 l \vert s'-s \vert}/6}
\bigg \{ 
   q^2 l^2 (s'-s)^2 
   \Bigl<{\rm e}^{i{\bf q}l{\tau}/3 \cdot \int_s^{s'} {\bf t}(s'\!') 
  {\rm d}s'\!'}\Bigr>' \qquad \qquad $$
$$
\qquad\qquad
-{{\partial} \over {\partial {\bf k}^2}} {\Big {\vert}}_{{\bf k}=0}
 \Big[ \Bigl<{\rm e}^{i({\bf k}+{\bf q})l{\tau}/3 
 \cdot \int_s^{s'} {\bf t}(s'\!') {\rm d}s'\!'}\Bigr>'
 - \Bigl<{\rm e}^{i{\bf k}l{\tau}/3 
 \cdot \int_s^{s'} {\bf t}(s'\!') {\rm d}s'\!'}\Bigr>'
 \Bigl<{\rm e}^{i{\bf q}l{\tau}/3 
 \cdot \int_s^{s'} {\bf t}(s'\!') {\rm d}s'\!'}\Bigr>'\Big]
 \bigg \} 
\eqno(2.19)
$$
where we have neglected numerical factors.  The prime in Eq. (2.19) 
indicates 
that the statistical average $\Bigl<{\rm e}^{i{\bf Q} \cdot \int_0^L
{\bf t}(s){\rm d}s} \Bigr>_{{\cal W}_{\rm t}} $ has been performed
using the factorization described in Eq. (2.18).  The evaluations of 
various quantities in Eq. (2.19) are given in Appendix A.

We now provide an argument that only the quantity arising from the
local interaction, i.e., from the terms with the structure
$\Bigl<{\rm e}^{i{\bf Q} \cdot \int_0^L {\bf t}(s){\rm d}s}
\Bigr>_{\rm rod}$ in Eq. (2.18), contribute to the persistence
length.  The long length scale term provides an interaction of the
excluded volume type and is not relevant in the calculation of $l_{\rm
p}$.  In order to show this let us split the integrals in Eq. (3.19)
into those arising from $|s'-s| < \eta$ and into the contributions
arising from scales larger than $\eta$.  The integrals involving the
tension term in ${\bf t}(s)$ are modulated by the Gaussian factor ${\rm
exp}(-q^2 l |s'-s|/6)$ which accounts for local fluctuations inside a
blob.  First consider the limit $q l \tau |s'-s| \ll 1$.  In this
limit we can approximate
$$
\left<{\rm exp}\left[{\mbox{$\frac{1}{3}$} i{\bf q}l{\tau} \cdot \int_s^{s'}
{\bf t}(s'') 
{\rm d}s''}\right] \right>_{{\cal W}_t} \approx  
{\rm exp}[-\mbox{$\frac{1}{54}$} q^2l^2{\tau}^2 (s'-s)^2]  
\eqno(2.20) 
$$
for $|s'-s| < \eta$.  It is clear from Eq. (2.20) that on scales less
than $S \ (\sim  1/l \tau^2)$ the Gaussian term is more important
(see also Eq. (2.7)) while on the scale $S < |s'-s| < \eta$ Eq. (2.20)
dominates.  On this length scale the segment of the chain behaves as a
rod, and hence charges separated by distance $|s'-s|= {\cal O}(\eta)$
can be assumed to interact via the screened Coulomb interaction.  Now
consider scales larger than $\eta$.  On these long length scales the
chain can be pictured as a sequence of rods of length $l_{\rm p} \sim
\eta l \tau$ surrounded by an impenetrable charge cloud of diameter
$d$.  Two such segments can come arbitrarily close together to a
distance of the order of $d$ because the chain is Gaussian on long
length scale.  The interaction of such rod segments gives rise to
 a short range
excluded volume type terms.  Since these interactions are not relevant for
describing the $local$ ordering (on scale $\eta$) we only include
contributions from short length scales in Eq. (2.19).  Polyions 
on length scale greater than $\eta$ can be
considered as flexible chains with an effective step length $l_{\rm
p}$ interacting via an effective excluded volume type interactions.

If we substitute Eqs. (A.5), (A.7), (A.8), and (A.11) into Eq. (2.29)
we obtain the result for
$l_{\rm p}$, which is valid for the screened polyions in the asymptotic
limit defined as $l_0 l_{\rm B} \ll A^2$,
$$
l_{\rm p} \sim \left({S_{} \over A}\right)^2 {l_{\rm B} \over D}
     \left(D + {1\over D} {1\over{\kappa}^2}\right),  \qquad\qquad   {\kappa} L \gg 1 
\eqno(2.21)
$$
where $(S_{}/A)^2 (l_{\rm B}/D) \sim {\cal O}(1)$ is the electrostatic 
energy of a blob.  The second term in Eq. (2.21), which varies as 
${\kappa}^{-2}$, can be identified as the electrostatic persistence
length while  
the $\kappa$-independent term ${\cal O}(1)D$ can be identified as a
renormalized bare persistence 
length due to the chain fluctuations.  
Surprisingly, this result is quite similar to the one derived from the OSF
result for
$l_{\rm p}$ by 
replacing $l_0, A$ by the blob size $D$ and the charge per monomer $e$ by
the charge $ge$ of a blob, as was done by Khokhlov and Khachaturian~\cite{KK}
(KK)
 without the benefit of
any explicit derivation:
$$
l_{\rm p}=l_0 + {l_{\rm B} \over 4 {\kappa}^2 A^2} \rightarrow  D + 
\left({S_{} \over A} \right)^2 {l_{\rm B} \over D}\cdot 
{1 \over D{\kappa}^2} $$
$$
\qquad\quad \ = D + {\cal O}(1) {1 \over D{\kappa}^2}
.\eqno(2.22)
$$
A derivation of the result displayed in Eq. (2.22) was also recently
provided by Li and
Witten [15] using an entirely different approach.  From the preceding
calculations, we find that the
fluctuations in chain conformations of asymptotic 
polyions $(l_0 l_B /A^2 \ll 1)$ do not  
invalidate the qualitative aspect of the OSF result.  They 
effectively reduce the direct
distances between charges so as to renormalize the bare parameters $l_0,
A$ into $D$ and $e$ into $ge$, as was first recognized by KK.   
  In 
this limit, local fluctuations inside the blobs are strong enough to
reduce 
significantly the direct distance between two consecutive charges by the 
factor $(l_0 l_{\rm B}/A^2)^{1/3}$ which is much less than 1.  This 
results in much stronger effective Coulomb repulsion between two consecutive
charges.  Since the length scale of $l_{\rm p}$ is much larger than 
the blob size $D$ and ordering as implied in these cases does not
refer to the local structure of 
 blobs, we can expect that the local fluctuations inside a blob
 not to affect significantly the property of the chain much beyond it.
  As a result we obtain qualitatively similar dependence of $l_{\rm p}$ 
  on $\kappa$ as that
obtained for stiff polyions. 
Note that this is relevant only for the asymptotic case where each blob
contains a large number of 
segments which roughly obey Gaussian statistics.  Indeed computer
simulations 
performed by Higgs and Orland~\cite{orland} correspond to the asymptotic case 
($l_0l_{\rm B}/A^2=0.05$), showing that $l_{\rm p}$ is much larger than 
${\kappa}^{-1}$.  Their results  support the expression for 
$l_{\rm p}$ given in 
Eq. (2.22).

The persistence length of unscreened polyions, however, diverges as
$L \rightarrow \infty$.  This is because all charges on the chain
contribute to the chain stiffening.  In the OSF theory it is proportional
to $l_{\rm B} L^2/A^2$.  To obtain the corresponding persistence length
for unscreened flexible polyions, we replace $l_{\rm B}/A^2$ by ${\cal
O}(1)/D$ and $L$ by $L'$
$$
l_{\rm p} \sim {{L'}^2 \over D}
.\eqno(2.23)
$$ 
It is interesting to contrast the results given in Eq. (2.22) with the
calculations of BJ.  Our theory and the BJ analysis use exactly the
same variational Hamiltonian.  The difference is that BJ identified
the persistence length by obtaining an upper bound to the free energy
of the system using the Gibbs-Bogoliubov inequality.  We, on the
other hand, have computed $\left< R^2 \right>$ using the ES method.
The persistence length of the chain is a  "local" measure of the size of
the polymer.  In the formulation of BJ, contributions from these
scales 
are averaged out whereas the ES method, at least approximately,
accounts for short length scales effects.  We believe this is
the major source for differing the conclusions drawn by BJ.  This
point is further illustrated in Appendix B where we show the
deficiencies of the variational theory, using the free energy for
computing "local" quantities such as persistence length.

\section{Polyions with $l_0 l_{\rm B}/A^2 \approx 
{\cal O}(1)$;
Gaussian Chain Model}

In the previous section we considered the case of flexible
polyelectrolyte in the so called asymptotic limit i.e., when $A^2 \gg
l_0 l_{\rm B}$.  If the magnitude of $A$, $l_0$, $l_{\rm B}$ are 
similar, then one is in the so called non-asymptotic limit and 
represents crossover
regime from small values to large values of $l_0 l_{\rm B}/A^2$.  In
the limit of $l_{0} l_{\rm B}/A^{2} \gg 1$, which is the best defined
limit, the electrostatic
persistence length is given by the Odjik-Skolnick-Fixman theory.  When
$l_0 l_{\rm B} /A^2 \sim 1$ then it is clear that (in all likelihood)
the blob size, $D$, is of the same order as $l_0$ or $D/l_0 \sim 1$.
In this limit the distinction between the blob size and the bare
persistence length is meaningless and the chain becomes ordered on
scales greater than $l_0$ due to electrostatic repulsion.  As a result
it seems reasonable that one can choose a simpler trial weight in
contrast to the polyions in the asymptotic limit which treated in the 
previous section.

Following the arguments given above, we chose the following trial
weight
for $l_0 l_{\rm B} /A^2 \sim 1$
$$
{\cal W}_1={\rm exp} \left[-{3 \over 2 l_1} \int \left(
{\partial{\bf r}(s) \over \partial s} \right)^2 \right]
\eqno(3.1)
$$ 
which is a Gaussian (non-interacting) theory with an effective step
length $l_1$.  The self-consistent
equation using the ES procedure, namely one that gives vanishing $\Delta R^2$ to first order in 
$$
B[{\bf r}(s)]={l_{\rm B} \over 2 A^2} \int\!\int {\rm d}s {\rm d}s'
{{\rm e}^{-\kappa |{\bf r}(s')-{\bf r}(s)|} \over |{\bf r}(s') -{\bf
r}(s)|} +\mbox{$\frac{3}{2}$} \left({1 \over  l}-{1 \over  l_1} \right)
\int_0^L \left({\partial
{\bf r} \over \partial s }\right)^2 {\rm d}s
,\eqno(3.2)
$$
for $l_1$ is 
$$
L l_1 \left({l_1 \over l}-1 \right)=\bigl<R^2 V \bigr>_{{\cal W}_1}-
\bigl<R^2 \bigr>_{{\cal W}_1} \bigl<V \bigr>_{{\cal W}_1}  
.\eqno(3.3)
$$
The calculation of the various quantities using the Gaussian weight is
straightforward, and using these results Eq. (3.3) becomes
$$
{1 \over l}-{1 \over l_1}={1 \over 18 L}{l_{\rm B} \over A^2}
\int\!\int {\rm d}s {\rm d}s' (s'-s)^2 \int { {\rm d}^3 q
\over (2 \pi)^3}{4 \pi q^2\over q^2+\kappa^2} {\rm e}^{-q^2 l_1 |s'-s|/6}
.\eqno(3.4)
$$
The evaluation of the integral in Eq. (3.4) is done by imposing a
lower cut-off for $q=q_0=l_1^{-1}$.  For this choice of $q_0$ the
contribution from charges separated by distance larger than $l_1$ is
exponentially small.  Furthermore, since $l_1$ is the only scale in the
Gaussian weight it is reasonable to set $q_0=l_1^{-1}$.  With this
cut-off the evaluation of the integrals in Eq. (3.4) is
straightforward and one obtains the following equation for $l_1$
$$
l_1^2 \left({1 \over l}-{1 \over l_1} \right)= {\alpha l_{\rm B}
\over A^2 \kappa^2}
\left\{1-\underline{{1 \over l_1 \kappa}
\left[{\pi \over 2}-{\rm tan}^{-1}\left({1 \over
l_1 \kappa}\right) \right]} \right\}
\eqno(3.5)
$$
where $\alpha$ is a numerical constant.

The self-consistent equation can be easily solved if $l_1$ is somewhat
larger than the Debye screening length $\kappa^{-1}$.  In this case,
we can neglect the underlined term in Eq. (3.5).  This results
in the following solution for $l_{\rm p}=l_1/2$
$$
l_{\rm p} \sim \cases{l_0+l_{\rm OSF},
\qquad \qquad &if $l_{\rm OSF} \ll l_0$
\cr {(l_0 l_{\rm B}/A^2)}^{1/2}{\kappa}^{-1},\qquad \qquad 
&if $l_{\rm OSF} \gg l_0$
\cr}
\eqno(3.6)
$$
where $l_{\rm OSF} \sim l_{\rm B}/A^2 \kappa^2$ is the electrostatic
persistence 
length of stiff chains derived from the variational equation.  
     For the flexible chains ($l_{\rm OSF} \ll l_0$ or
equivalently $l_1 \gg l_0$), the
persistence length  varies as ${\kappa}^{-1}$, which is in accord with
Barrat and Joanny [13] and our earlier calculation [10].  The result in
Eq. (3.6) has also been obtained by others using different methods~\cite{liu}.  
These two distinct
scaling regions are obtained according to
whether the bare persistence length $l_0$ is larger or smaller than 
$l_{\rm OSF}$. The dependence of $l_1$ on $l_0$
for the flexible chains ensures the {\it cross-over}; namely the value of
the 
electrostatic persistence length for the flexible chain limit crosses
over to the  
Odjik result at $l_0 \simeq l_{\rm OSF}$.

The result for stiff chains ($l_{\rm OSF} \ll l_0$) is consistent with
the assumption that $l_1 > \kappa$.  Thus this is valid for wide range of
parameters as long as the inequality $l_{\rm OSF} \ll l_0$ holds.  On
the other hand, it is not clear if we can neglect the underlined term
in Eq. (3.5) for intrinsically flexible chains.   If we include the
contribution
from the underlined term,   
$l_{\rm p}$ may not show simple scaling law exhibited in Eq. (3.6).
As long as $l_{\rm p} \gg 
l_0$, however,  the self-consistent equation (3.5) admits a solution
for which $l_{\rm
p}$ varies as $\kappa^{-1}$.   This
can be easily seen if we rewrite Eq. (3.5) in terms of
$x=l_1 \kappa$ as follows 
$$
x^2  \sim {l_{\rm B}l_0 \over A^2} 
\left\{1-{{1 \over x}
\left[{\pi \over 2}-{\rm tan}^{-1}\left({1 \over
x} \right) \right]} \right\}, \qquad l_{\rm p} \gg l_0
.\eqno(3.7)
$$
This is a transcendental equation for $x$ which cannot be solved
analytically.  The solution for $x$, however, is a function of $l_0 l_{\rm
B}/A^2$ only which does not assume a simple form except for a case of $l_1
\kappa \gg 1$.  In general, we can write
$$
l_{\rm p} = {\cal G}\left({l_0 l_{\rm B} \over A^2}\right) \ \kappa^{-1},
\qquad l_{\rm p} \gg l_0
\eqno(3.8)
$$
where ${\cal G}$ is a function of $l_0 l_{\rm B}/ A^2$ only.
Thus, the behavior of the persistence length which varies as $\kappa^{-1}$
is the
general rule for the case of $l_{\rm p} \gg l_0$ even though its
dependence on $l_0 l_{\rm B}/A^2$ is complicated.  Explicitly ${\cal
G}$ satisfies 
the following equation
$$
(2{\cal G})^2 \sim {l_{\rm B}l_0 \over A^2} 
\left\{1-{{1 \over 2{\cal G}}
\left[{\pi \over 2}-{\rm tan}^{-1}\left({1 \over
2{\cal G}} \right) \right]} \right\}
,\eqno(3.9)
$$     
which can be solved graphically.  As mentioned earlier, the above equation
allows a simple solution, i.e, ${\cal G} \sim (l_0 l_{\rm B}/A^2)^{1/2}$,
if
${\cal G} \gg 1$ or $l_{\rm p} \gg \kappa^{-1}$.  Otherwise 
we expect a different functional behavior of ${\cal G}$.    

According to the above analysis,  the persistence length of polyions
in the non-asymptotic limit, which is assumed to be well described by a
trial Gaussian weight with an effective step length $l_1$, 
shows a
simple scaling behavior only when
$l_{\rm p} \gg l_0$.  On the other hand it is not clear if this is valid
for other range of parameters.  In order to test this we solved the
self-consistent equation (3.38) numerically for a few cases, with
$\alpha l l_B/A^2$ ranging from 2-6.  The values of $A$,
$l_{\rm B}$,  and $l$ were chosen to be unity.
All lengths are estimated in unit of $l=1$.  In  Fig. 1, ${\rm
ln} \ l_{\rm e} $ is 
plotted against ${\rm ln} \ \kappa^{-1}$ .
This figure shows that 
when  $\alpha l_{\rm B}l/A^2$ exceeds 3 then $l_{\rm e}$ has a sub-linear
dependence on $\kappa^{-1}$, i.e., $l_{\rm e} \sim
\kappa^{-y}$ with $y$ varying from 0.6-0.9.  These results appear to be in accord with the recent
simulations
performed by Micka and Kremer~\cite{micka}.    The experimental data of S. F${\ddot {\rm o}}$rster et al.~\cite{forster} also suggest that the persistence length does not show a
simple power law behavior in certain regimes.  When $\alpha l_Bl/A^2$
is less than about 3 there is no unique scaling behavior, i.e., $l_e$
does not vary as $\kappa^{-y}$.   

\section{Crossover Conditions and Comparison with Simulations}

Based on 
Monte Carlo simulations for weakly charged polyelectrolytes, Micka and
(MK)~\cite{micka} have arrived at few interesting conclusions.  They
varied both $\kappa$ and the intrinsic stiffness of the chain.  Their
principle findings are: (a) In the limit of large stiffness, the
electrostatic persistence length $l_e \sim \kappa^{-2}$ as predicted
by the OSF theory.  This is valid when the intrinsic persistence
length exceeds the electrostatic persistence length, i.e., $l_0 \gg
l_e$.  For flexible chains MK find that $l_e \sim \kappa^{-y} (y <
1)$, and $y \rightarrow 1$ as $l_0 \approx l_e$.  (2) MK also showed
that $l_p$ is a strong 
function of the intrinsic persistence length, $l_0$.  This implies
that $l_0$ is altered by balance between the bending rigidity and the
electrostatic interactions.  They suggest that the contour distance,
$s_c$, at which the contributions from the bending 
rigidity and electrostatic interactions 
become comparable goes as $s_c \sim \kappa^x$
with $x \ge 0$.

Our crossover condition for the change in the persistence length goes
from the OSF result $(y=2)$ to $y=1$ is predicted to occur when
$l_{OSF} \approx l_0$ (See Eq. (3.b)).  We will argue that this is not
inconsistent with the findings of MK.  According to Eq. (3.6), when
$l_{OSF} \approx l_0$, the persistence length $l_p \approx l_e \approx
\sqrt{l_{OSF} l_0} \approx  \kappa^{-1}$.  This is the
crossover condition for the transition from $y=1$ to $y=2$ reported by
MK.

In their paper, MK suggested that the $\kappa$-dependence of the
contour crossover
 length from our earlier theory gives $s_c \sim \kappa^{-2}$.  We believe
this is an erroneous interpretation of our earlier theory (See
Eq. (15) of Ref.~\cite{thirum.ma}).  The crossover length for
deviation from OSF behavior for $l_e$ need not be the same as $s_c$, which
describes the contour length scale when electrostatic effects become
comparable to rigidity.  BJ computed $s_c$ by considering $\left<
\theta^2 (s) \right>$ where $\cos \theta (s) ={\bf u}(s) \cdot {\bf u}(0)$.
 When $\left< \theta^2 (s) \right>$ varies as $1/(l_0+l_{OSF})$, we
expect that the bending term is dominant.  By identifying the crossover
length 
scale
$s_c$ to be a length scale where the value of $\left< \theta^2 (s) \right>$ crosses
over from $1/l_0$  to $1/(l_0+l_{OSF})$, BJ showed that $s_c \sim
\kappa^{-1}$.  

We can compute $s_c$ directly using an variational equation for
flexible chains when $l_0l_B/A^2 \approx {\cal O}(1)$, i.e., in the
so-called non-asymptotic limit.  The contour crossover length can be
computed by analyzing Eq. (3.4).  The
condition for computing $s_c$ is that $l_e \approx l_0$, which implies
using Eq. (3.4)
$$ {1 \over l_0} \approx {1 \over 18} {l_B \over A^2} \int_0^{s_c} ds
s^2 \int {d^3 q \over (2 \pi)^3} {4 \pi q^2 \over q^2+\kappa^2}{\rm
e}^{-q^2 l_1 s/6}
.\eqno(4.1) 
$$
This equation  is too complicated to solve analytically.  However, it is easy to
solve Eq. (4.1) numerically. The resulting solution shows that $s_c$ is
an increasing function of $\kappa$. The increase is not very dramatic which
suggests that $s_c \sim\kappa^{x}$ with ${x}$ being small. The parameters used in
solving Eq. (4.1) are l = A = $l_B$ = 1. At least
in this case our results are consistent with the numerical results of MK.  A
more detailed comparison between Eq. (4.1) and the simulations would
require a thorough  mapping of the variables in the theory and the
simulation parameters.  We believe that, at least qualitatively, the
theory is in accord with the computer simulation results.  

\section{Conclusions}

The variation of the electrostatic persistence length of flexible
charged polyelectrolytes with $\kappa$ is complicated.  Several regimes
can emerge depending on the controlling parameter $l_0 l_B/A^2$. Here
we have inentified the limiting behavior in certain regions of the
parameter space.
The computation of the electrostatic persistence length for flexible
chains  in the asymptotic limit $(l_0 l_B /A^2 \ll 1)$ suggests that
the original OSF result  is valid.  These results were obtained by a
trial Hamiltonian mimicking a semiflexible chain under
tension~\cite{joanny}  
together with the uniform expansion method~\cite{es}.  It appears that by
focusing directly on the calculation of the mean square end-to-end
distance effects that are relevant on the length scale on the order of
 persistence length can be implicitly taken into account using
Edwards-Singh method.  The results presented
here together with our earlier theory~\cite{thirum.ma} give a rather complicated
picture of the dependence of the electrostatic persistence length on
the inverse Debye screening length.  We predict that in the regime
when $l_0 l_B/A^2$ is very large or very small the essential features
of the OSF result is valid.  In the limit of small $l_0 l_B/A^2$ the
blob picture suggested by Khokhlov and Khachaturian~\cite{KK} appears to be
valid.  In this case, the mean distance between the charges is roughly
a $\kappa$-independent blob size with each blob carrying a renormalized
 charge.  In the non-asymptotic regime the scaling of $l_e$ with
$\kappa$ is more complicated.  Here we can expect fractional power for
the dependence of $l_e$ on $\kappa$ or in some instances complete
absence of simple scaling laws.  Since $l_0$ and $A$ can be
experimentally varied over a large regime it may not be possible to
analyze the data on the dependence of the persistence length of 
polyelectrolytes using simple scaling laws (Cf. Beer et
al.~\cite{beer}). \\

{\bf Acknowledgements}: This work was carried out by B. -Y. Ha to
satisfy the requirements for a Doctorate degree at the University
of Maryland. We are grateful to the National Science Foundation (through
grant number CHE96-29845) for support.

\noindent
{\bf Appendix A}\\

In this appendix, we describe the evaluation of the quantities 
appearing in Eq. (3.18).  We have argued in the text of the
paper that the contributions on large length scale (greater than 
$\eta$) induce an interaction of
the excluded volume type, and does not affect the local ordering which 
is described in terms of a persistence length.  
The average of $\bigl<{\rm exp}[i({\bf k}+{\bf q})l \tau /3
\cdot \int_s^{s'} {\bf t}(s''){\rm d}s''] \bigr>'$ on the short length
scale can be approximated as
$$
{l_{\rm B} \over A^2}{1\over L'}\int\!\int_{|s'-s| \le \eta}
\Bigl[- {\partial \over \partial {\bf k}^2} \Bigr]_{{\bf k}=0}
\int { {\rm d}^3q \over ({\bf q}-{\bf k})^2+\kappa^2}{\rm e}^{-({\bf
q}-{\bf k})^2 l
|s'-s|/6} 
\Bigl< {\rm e}^{i{\bf q}l \tau/3 \cdot \int_s^{s'}{\bf t}(s''){\rm
d}s''}\Bigr>' $$
$$
\approx I_1 + I_2 + I_3
\eqno(A.1)$$
where
$$I_1 = 
{l_{\rm B} \over A^2}{1 \over l \tau}\int_0^{S_{}}{\rm
d}s \Bigl[ -{\partial \over \partial \kappa^2} \Bigr]
\int {{\rm d}^3q \over q^2 +\kappa^2} {\rm e}^{-q^2 l s/6}
\eqno(A.2)
$$
$$
I_2=
{l_{\rm B} \over A^2}{l \over  \tau}\int_0^{S_{}}{\rm
d}s s^2
\int {q^2{\rm d}^3q \over q^2 +\kappa^2} {\rm e}^{-q^2 l
s/6}\eqno(A.3)
$$
$$
I_3={l_{\rm B} \over A^2}{1 \over L'}
 \int\!\int_{{S_{}}\le |s'-s| \le \eta} {\rm d}s{\rm d}s'
\Bigl[- {\partial \over \partial \kappa^2} \Bigr]
\int { {\rm d}^3q \over q^2 +\kappa^2} 
\Bigl< {\rm e}^{i{\bf q}l \tau/3 \cdot \int_s^{s'}{\bf t}(s''){\rm
d}s''}\Bigr>_{\rm rod} \Bigr\}
.\eqno(A.4)
$$
The first term in Eq. (A.1), $I_1$ can be expanded in powers of
$\kappa^2 S l$ (which is much smaller than 1)
$$
I_1={l_{\rm B} \over A^2} {D \over l^2}
\Bigl[ -{\partial \over \partial \kappa^2} \Bigr]
{1 \over \kappa} \Bigl\{ 1-\bigl[1-\Phi(\kappa \sqrt{l S_{}/6})\bigr]
{\rm e}^{\kappa^2lS_{}} \Bigr\} 
$$
$$
\approx(S_{}A)^2 l_{\rm B}[1+{\cal
O}(\kappa^2l S_{})] \qquad\qquad\qquad\quad\quad
\eqno(A.5)
$$
where $\Phi(x)$ is the error function defined by
$$\Phi(x)=
{2 \over \sqrt{\pi}} \int_0^x {\rm e}^{-t^2}{\rm d}t
.
\eqno(A.6)
$$
Similarly , $I_2$, (Eq. (A.2)) can also be computed.  In this case we set
$\kappa$ to zero right at the outset without any loss of generality.
Since this integration involves length scale which is much smaller
than $\kappa^{-1}$, the interaction between charges within this length
scale can be assumed to be unscreened.  Thus $I_2$ is approximately
given by
$$I_2=
{l_{\rm B} \over A^2}{l \over \tau}\int_0^{S_{}} {\rm d}s s^2
\int {\rm d}^3 q {\rm e}^{-q^2 l s/6}
\approx (S_{}/ A)^2 l_{\rm B}
.\eqno(A.7)
$$

Finally $I_3$ can be approximated as 
$$
I_3=
{l_{\rm B} \over A^2}{D^2 \over l^2} 
\Bigl[-{\partial \over \partial \kappa^2} \Bigr] 
{1\over L'}\int\!\int_{D \le |x'-x| \le l_{\rm p}}
{\rm d}x {\rm d}x' \int {{\rm d}^3 q \over q^2+ \kappa^2} 
\Bigl<{\rm e}^{i {\bf q} \cdot \int_x^{x'}{\bf t}(x''){\rm d}x''}
\Bigr>_{\rm rod} $$
$$
={l_{\rm B} \over A^2}{D^2\over l^2}
\Bigl[-{\partial \over \partial \kappa^2} \Bigr] 
\int_D^{l_{\rm p}} {\rm d}x {{\rm e}^{-\kappa x} \over
x}\qquad\qquad\qquad\qquad\qquad\qquad\qquad\quad
$$
$$
 \approx \Bigl({S_{} \over A}\Bigr)^2 {l_{\rm B}\over \kappa^2}
\Bigl[1+ {\cal O}(\kappa D)+{\cal O}({\rm e}^{-\kappa l_{\rm
p}})\Bigr]
 \qquad\qquad\qquad\qquad\quad\qquad \ \eqno(A.8)
$$
By following exactly the same procedure the first term in Eq. (3.19)
can be approximated as sum of two terms, namely, 
$$
{l_{\rm B} \over A^2 }{l^2 \over L'}
\int\!\int_{S_{} \le |s'-s| \le \eta}
\int {{\rm d}^3 q \over q^2 + \kappa^2} q^2 (s'-s)^2 {\rm e}^{-q^2 l
|s'-s|/6} \Bigl< {\rm e}^{i
{\bf q} l \tau /3 \cdot \int_s^{s'} {\bf t}(s''){\rm d}s''} \Bigr>'
=I_4+I_5
\eqno(A.9)
$$
where $I_4$ is the contribution to Eq. (A.9) coming from the scale
$|s'-s| < S$ and is given by 
$$
I_4 \approx l_{\rm B} {S^2 \over A^2}
.\eqno(A.10)
$$
The term $I_5$ arises from contributions on the large length scale $S
\le |s'-s| \le \eta$ and can be approximated as followed
$$
I_5 \approx
{l_{\rm B} \over A^2 }{D^4 \over l^2} {1 \over L'} \int\!\int_{ D \le
|x'-x| \le
l_{\rm p}}(x'-x)^2 \int {q^2 {\rm d}^3q \over q^2 +\kappa^2} 
\Bigl<{\rm e}^{-i{\bf q} \cdot \int_x^{x'} {\bf t}(x'') {\rm
d}x''}\Bigr>_{\rm rod} 
$$
$$
\approx {l_{\rm B} \over A^2 }{D^4 \over l^2}\int_D^{l_{\rm p}} {\rm d} x
x^2
{ \partial^2 
\over \partial x^2} {{\rm e}^{-\kappa x} \over x}
\qquad\qquad\qquad\qquad\qquad\qquad\qquad \ 
$$
$$
\approx l_{\rm B} S_{}^2/A^2 \ \ \
\qquad\qquad\qquad\qquad\qquad\qquad\qquad\qquad\qquad\qquad
\eqno(A.11)
$$ \\

\noindent
{\bf APENDIX B} \\

To illustrate the limitation of the Gibbs-Bogoliubov variational 
approach in the computation of persistence length, let us consider a simplified 
system which is described by the following Hamiltonian
$$
{{\cal H} \over k_{B}T}= {l_{0} \over 2} \int_{0}^{L} 
\dot{\theta}^{2} (s) {\rm d}s + \mbox{$\frac{1}{2}$} \int_{0}^{L}
\int_{0}^{L}
\dot{\theta}(s) 
K(s,s') \dot{\theta}(s'){\rm d}s {\rm d} s'
\eqno(B.1)
$$
where $\cos \theta (s)={\bf u}(s) \cdot {\bf u}(0)$ and  $\dot{\theta 
}(s)={ \partial \theta (s) \over \partial s}$.  The kernel 
$K(s,s')$ describes interaction between charges on the chain and has 
yet to be specified.  If the kernel $K(s,s')$ is 
$$K(s,s')= {1 \over A^{2}} \int_{0}^{s'+{L \over 2}} ds_{1} \int_{s+{L
\over 
2}} { dv \over ds} \bigg|_{s_{2}-s_{1}} { (s_{2}-s)(s'-s_{1}) 
\over (s_{2}-s_{1})},\eqno(B.2)
$$
 then the Hamiltonian in Eq. (B.1) corresponds to a polyelectrolyte 
chain near the rod limit.  For the present purpose, we use a 
simpler form of $K(s,s')$ whose Fourier transform is given by
$ K_{q}={ \kappa^{2} \over \kappa^{2}+2 q^{2}} l_{OSF}$:
$$ {{\cal H} \over k_{B}T} \approx \mbox{$\frac{1}{2}$} 
 \sum_{q} \theta_{q} \theta_{-q}\,\Bigl[l_{0} + K_{q} \Bigr] 
 .\eqno(B.3)
 $$
Since the Hamiltonian in Eq. (B.3) is quadratic in the variables
$\theta_q$ all quantities with reference to Eq. (B.3) may be exactly
computed.  The efficacy of the approximate methods can be assessed by
comparing to the exact result.  In the following, we adopt 
three different trail Hamiltonians in the computation of the persistence 
length to illustrate possible limitations in each case. 

(a) First, let us use a trial Hamiltonian
$$ {{\cal H}_{t} \over k_{B}T}=\mbox{$\frac{1}{2}$}  \sum_{q} 
q^{2} \tilde{K}_{q} \, \theta_{q} \theta_{-q}
.\eqno(B.4)$$   
Within the Gibbs-Bogoliubov variational approach, the best choice for
 $\tilde{K}_{q}$ is obtained by minimizing the variational 
free energy $ {\cal F}_{v}={\cal F}_{t}+ \bigl< {\cal H}-{\cal 
H}_{t} \bigr>_{t} $.  Performing the Gaussian integration over $\theta 
(s)$, we have
$$ {{\cal F}_{v} \over k_{B}T}= \mbox{$\frac{1}{2}$} \sum_{q} 
\log \left( { q^{2} 
\tilde{K}_{q} \over \pi }\right) + \mbox{$\frac{1}{2}$} 
\sum_{q}\left( { l_{0}+K_{q} \over \tilde{K}_{q}}-1 \right)
.\eqno(B.5)$$ 
Minimization of the variational free energy with respect to
$\tilde{K}_{q}$ 
leads to
$$
\tilde{K}_{q}= l_{0} + l_{OSF}{ \kappa^{2} \over \kappa^{2} + 2 
q^{2}}.\eqno(B.6)$$
When a general form of variational kernel is used, the variational 
model produces the same result as the original Hamiltonian.

(b) As a second example, let us choose a bending model as a trial 
system which is described by
$$ {\cal H}_{t} ={ l_{p} \over 2} \sum_{q} q^{2} \theta_{q}\,
\theta_{-q}.\eqno(B.7)$$
This is such that $\bigl< \theta^{2} (s) \bigr> = s \big/ l_{p}$.  In 
this case, the variational free energy is given by, neglecting 
numerical prefactors,
$$ {{\cal F}_{v} \over k_{B}T} \sim {L \over \Lambda} {l_{0} \over 
l_{p}}
+  { L \over l_{p}} \kappa l_{OSF} \eqno(B.8)
$$ 
where $\Lambda^{-1}$ is the upper bound for $q$.  Minimization of 
the variational free energy results in 
$$ l_{p} \sim l_{0} + {\Lambda l_{B} \over A^{2}} { 1 \over 
\kappa}
.\eqno(B.9)$$ 
When a bending model is used, the Gibbs-Bogoliubov variational approach
leads to a wrong result for the electrostatic persistence 
length of a polyelectrolyte chain near the rod-limit which scales as 
$\kappa^{-1}$.  This is because the free energy of the bending model 
cannot be made mimic the free energy on scales on the order of the
persistence length.  In other words, short length 
scale ($ l_{p}$)
properties in the original system are washed out by minimizing the 
variational free energy with respect to $l_{p}$.

(c) Let us now compute the persistence length of a polyelectrolyte 
chain described by ${\cal H}$ in Eq. (B.1) using the Edward-Singh's 
self-consistent method.  The following derivation is analogous 
to that given in the text (Cf. Eq. (3.11)).  To this end, we write
$$
{{\cal H} \over k_{B}T} = {l_{p} \over 2} \sum_{q} q^{2} 
\theta_{q} \theta_{-q} + \underline{ \left( {l_{0} \over 2}-{l_{p} 
\over 2} \right) \sum_{q} q^{2} \theta_{q} \,  \theta_{-q}+  
\mbox{$\frac{1}{2}$} \sum_{q} K_{q} q^{2}\theta_{q}\, \theta_{-q}}
\eqno(B.10)$$
and compute $\big< \theta^{2}(L) \bigr>$ self-consistently by 
considering the underlined term to be small.  Up to the first order in 
the underlined term, the bending model can be made the same as the 
real system by requiring
$$
(l_{0}-l_{p}) { \partial \over \partial l_{p}} \bigl< 
\theta^{2}(L) \bigr>_{t} = \bigl< \theta^{2}(L)  \mbox{$\frac{1}{2}$} 
\sum_{q} K_{q} q^{2} \theta_{q}\, \theta_{-q} \bigr>_{t}- 
\bigl< \theta^{2}(L) \bigr>_{t} \bigl<\mbox{$\frac{1}{2}$} 
\sum_{q} K_{q} q^{2} \theta_{q}\, \theta_{-q} \bigr>_{t}
.\eqno(B.11)$$
A little algebra leads to the following equality from the Eq. (B.11)
$$(l_{0}-l_{p}) { \partial \over \partial l_{p}} \bigl< 
\theta^{2}(L) \bigr>={2 \over l_{p}^{2}} \sum_{q} K_{q} { 
\sin^{2} (q L/2) \over q^{2}}
.\eqno(B.12)$$
For large $L$ most of contribution to the sum on r.h.s of Eq. (B.12) is 
from $q \simeq 0$.  This results in
$$
l_{p} \approx l_{0}+ l_{OSF}
.\eqno(B.13)$$
Thus the ES method provides a reliable way of calculating the
persistence length $l_p$.  The result using this method agrees with
the exact answer for the simple Hamiltonians of the form given in
Eq. (B.3). 
 By
focusing directly on the local quantity the ES method implicitly
includes effects on the scale of the persistence length.  Such fluctuations
are integrated out when the free energy is computed.  Consequently,
the persistence length needs not be accurately given by minimization
of an approximate free energy.

\newpage
\noindent
{\bf Figure Caption}\\

\noindent
Fig. (1).  Plot of the electrostatic persistence length, $l_e$, as a
function of $\kappa^{-1}$.  Both quantities are expressed in natural
logarithm values.  The various curves correspond to different values
of the parameter $\alpha l_B l/A^2$ (See Eq. (3.5)). The values of
$\alpha$ from the top to the bottom curves are 6, 4, 3, 2.5, and 2
respectively.
For $\alpha l_B l/A^2 \ge 2.5$, $l_e \propto
\kappa^{-y}$ and the exponent can be obtained from the slope of the
displayed curves.  The exponent y varies  from 0.6-0.9.  For $\alpha l_B l/A^2
\le 2.5  $, there appears to be no
simple power law describing the dependence of $l_e$ on $\kappa$.  

\end{document}